\documentclass[pra,twocolumn,showpacs]{revtex4}
\usepackage{graphicx}
\begin{document}

\bibliographystyle{unsrt}

%%%%%%%%%%%%%%%%%%%%%%%%%%%%%%%%%%%%%%%%%
\title{The role of pump coherence in two-photon interferometry}
\author{J. Liang, S.M. Hendrickson and T.B. Pittman}
\affiliation{Physics Department, University of Maryland Baltimore
County, Baltimore, MD 21250}
%%%%%%%%%%%%%%%%%%%%%%%%%%%%%%%%%%%%%%%%%

%%%%%%%%%%%%%%%%%%%%%%%%%%%%%%%%%%%%%%%%%
\begin{abstract}
We use a parametric down-conversion source pumped by a short
coherence-length continuous-wave (CW) diode laser to perform
two-photon interferometry in an intermediate regime between the more
familiar Franson-type experiments with a long coherence-length pump
laser, and the short pulsed pump ``time-bin'' experiments pioneered
by Gisin's group. The use of a time-bin-like Mach-Zehnder
interferometer in the CW pumping beam induces coherence between
certain two-photon amplitudes, while the CW nature of the experiment
prevents the elimination of remaining incoherent ones. The
experimental results highlight the role of pump coherence in
two-photon interferometry.
\end{abstract}
%%%%%%%%%%%%%%%%%%%%%%%%%%%%%%%%%%%%%%%%%


\maketitle

\section{Introduction}
\label{sec:introduction} \vspace*{-.1in}

Two-photon interferometry is a rapidly growing field with
applications ranging from metrology \cite{dangelo01,giovenetti04}
and secure communications \cite{ekert92} to fundamental tests of
quantum mechanics \cite{greenberger93}. When the two-photon state is
produced by parametric down-conversion (PDC) \cite{klyshkobook}, it
is well-known that the coherence properties of the laser pumping the
PDC source play an important role. The original Franson
interferometer experiments \cite{franson89} represent one extreme,
where a long coherence-length continuous-wave (CW) pump laser is
used to produce a PDC photon-pair whose emission time is essentially
undefined. In the other extreme, Gisin's time-bin entanglement
experiments \cite{brendel99} utilize a single short pump pulse
passing through an unbalanced Mach-Zehnder (MZ) interferometer to
produce a PDC photon-pair whose emission time is a coherent
superposition of two well-defined values.

In this paper we provide an explicit experimental demonstration of
an intermediate regime, where a {\em short coherence-length CW pump
laser} is passed through a Gisin-type MZ pump interferometer. This
produces a PDC photon-pair in a coherent superposition of two time
values with a well-defined separation, but an unknown overall
starting time. The coherence generated between two distinct creation
times allows a recovery of certain two-photon interference effects
that would otherwise be non-existent with a short coherence length
CW pump, while the incoherent nature of the possible starting times
limits the maximum visibility of these interference effects to
$50\%$.

This work builds on several earlier PDC experiments in which
two-photon interference effects originate from the possibility of
multiple arrival times of a given pump photon at the PDC source. For
example, Burlakov  {\em et.al.} used a CW Helium-Cadmium laser
possessing several randomly-phased longitudinal modes which
interfere to produce a series of repetitive spikes in the pump
photon amplitude \cite{burlakov01}. Similar effects were observed by
Kwon {\em et.al.} using a low-cost multimode diode laser as a pump
\cite{kwon09}. Other related work involves the coherence between two
subsequent pump pulses from a mode-locked laser
\cite{keller97,kim99,kim00}, and higher order time-bin entanglement
generated by a coherent pulse train \cite{deriedmatten04,stucki05}.

\begin{figure}[b]
\includegraphics[width=2.2in,angle=90]{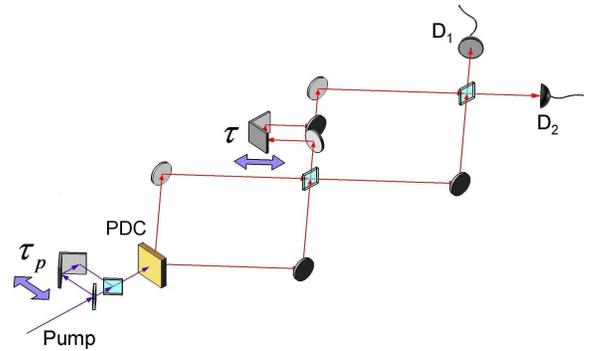}
\caption{Overview of the parametric down-conversion (PDC) experiment
used for two-photon interferometry in an intermediate regime between
the original Franson-type experiments with a narrowband
continuous-wave (CW) pump laser \protect{\cite{franson89}}, and the
Gisin time-bin type experiments with a short pulse as a pump
\protect{\cite{brendel99}}. Here, the PDC source is pumped by a
short-coherence length CW laser, and two sequential unbalanced
Mach-Zehnder interferometers are used to experimentally demonstrate
the role of pump coherence.}
\end{figure}

\section{Overview}
\label{sec:overview} \vspace*{-.1in}

A conceptual overview of the experiment is shown in Figure 1.  A PDC
source emits photons into opposite ports of a 50/50 beamsplitter in
a typical Hong-Ou-Mandel (HOM) interferometer arrangement
\cite{hong87,shih88}. The HOM interference effect causes both
photons to leave in the same output port, which produces a
two-photon N00N state \cite{boto00} in the output modes. The outputs
of the HOM beamsplitter are then recombined at a second 50/50
beamsplitter, thereby forming a PDC MZ interferometer of the kind
first developed by Rarity {\em et.al.} \cite{rarity90}. In order to
demonstrate the effects of interest here, we modify Rarity's
balanced MZ interferometer by including a very large path length
imbalance, denoted by $\tau$ in Figure 1. The value of $\tau$ is
chosen to be much greater than the coherence times of PDC photons as
well as the pump laser. The setup is completed by the inclusion of a
Gisin-type MZ pump interferometer, whose path-length imbalance
$\tau_{p}$ is carefully matched to $\tau$. The experiment is
conducted by recording the coincidence counting rate between
detectors $D_{1}$ and $D_{2}$ as a function of the relative phase
difference of the interferometers.

Typical Franson-type or time-bin entanglement experiments require
two spatially separated PDC MZ interferometers to study nonlocal
correlations of distant photons
\cite{franson89,franson91,marcikic04}. However, many key features of
the role of pump coherence in two-photon interferometry can be
explored in a less stringent setting using only one common PDC MZ
interferometer. In many cases, the use of a single interferometer
can simplify the experiments, while still allowing the essential
physics to be demonstrated. This idea has been used in some of the
earliest experimental work on two-photon interferometry
\cite{kwiat90,brendel91}.

For our purposes, the use of a two-photon N00N state in the single
unbalanced PDC MZ interferometer of Figure 1 simplifies the
experimental requirements in two ways. First, it reduces the number
of unbalanced MZ interferometers which must be experimentally
stabilized and phase-controlled. Second, the photon-bunched form of
the N00N state eliminates the need to electronically ``cut-off''
joint detection events corresponding to distinguishable two-photon
amplitudes associated with one PDC photon traveling a long path
while its twin takes a short path through an unbalanced MZ
interferometer. For a given PDC source, this effectively doubles the
coincidence counting rate compared to the case of two separately
seated PDC MZ interferometers. As can be seen from Figure 1, when
$\tau_{p} = \tau$ this leaves an output state with exactly four
indistinguishable terms contributing to a coincidence count between
$D_{1}$ and $D_{2}$ at some time $t$:

\begin{eqnarray}
|\Psi\rangle &=& \frac{1}{2} \{ e^{i[\phi_{o}(t^{\prime}-\tau)+
\phi]} |S_{p}; L_{s}L_{i}\rangle + \nonumber \\
&&e^{i[\phi_{o}(t^{\prime}-\tau) + \phi_{p}]}|L_{p};
S_{s}S_{i}\rangle \ + \nonumber \\
&&e^{i[\phi_{o}(t^{\prime})]} |S_{p}; S_{s}S_{i}\rangle + \nonumber \\
&&e^{i[\phi_{o}(t^{\prime}-2\tau)+ \phi_{p} + \phi]}|L_{p};
L_{s}L_{i}\rangle \} \label{eq:4terms}
\end{eqnarray}

In this simplified notation, the first ket $|S_{p};
L_{s}L_{i}\rangle$ denotes the pump photon traveling through the
short arm of the pump interferometer, while the signal and idler
photons of the PDC pair take the long path through the PDC MZ
interferometer. In analogy, the final three kets in equation
(\ref{eq:4terms}) correspond to the three other possible
combinations of long and short paths through the two
interferometers.

Each of the four indistinguishable amplitudes in equation
(\ref{eq:4terms}) has a different relative phase. For simplicity, we
use $\phi_{p}$ to represent the relative phase difference of the
long vs. short arm in the pump interferometer. Analogously, $\phi$
corresponds to the ``long-short'' relative phase difference in the
PDC MZ interferometer. In our setup, the phase $\phi$ represents the
accumulated relative phase difference of a two-photon N00N state in
the PDC MZ interferometer, with a period corresponding to the
biphoton deBroglie wavelength
\cite{rarity90,jacobson95,fonseca99,edamatsu02}.

The term $\phi_{o}$ describes the initial phase of the pump laser.
Because we are using a short coherence-length CW pump, the time
dependence of $\phi_{o}$ plays a particularly important role.  For a
coincidence detection event at time $t$, we use
$\phi_{o}(t^{\prime})$ to denote the value of the pump laser phase
at an earlier time $t^{\prime}= t-x/c$, where x is the overall
optical path length through the device via the short arms of both MZ
interferometers.  The first and second terms in equation
(\ref{eq:4terms}) correspond to an overall extra delay of $\tau$
(eg. propagation through the long arm of only one MZ interferometer)
and therefore contain the initial pump phase
$\phi_{o}(t^{\prime}-\tau)$. Likewise, the fourth term corresponds
to propagation through the long arm of both MZ interferometers and
thus contains the initial pump phase at an even earlier time,
$\phi_{o}(t^{\prime}-2\tau)$.

Because $c \tau$ is intentionally set to be much greater than the
coherence length of the pump laser, the value of
$\phi_{o}(t^{\prime}-\tau)$ will have no definite relation to
$\phi_{o}(t^{\prime})$ (or $\phi_{o}(t^{\prime}-2\tau)$).
Consequently, the first and second terms in equation
(\ref{eq:4terms}) can interfere \cite{brendel99}, but the third and
fourth terms provide an incoherent background which limits the
visibility of the two-photon interference effects.

\begin{figure*}
\centering
\includegraphics[width=5.5 in]{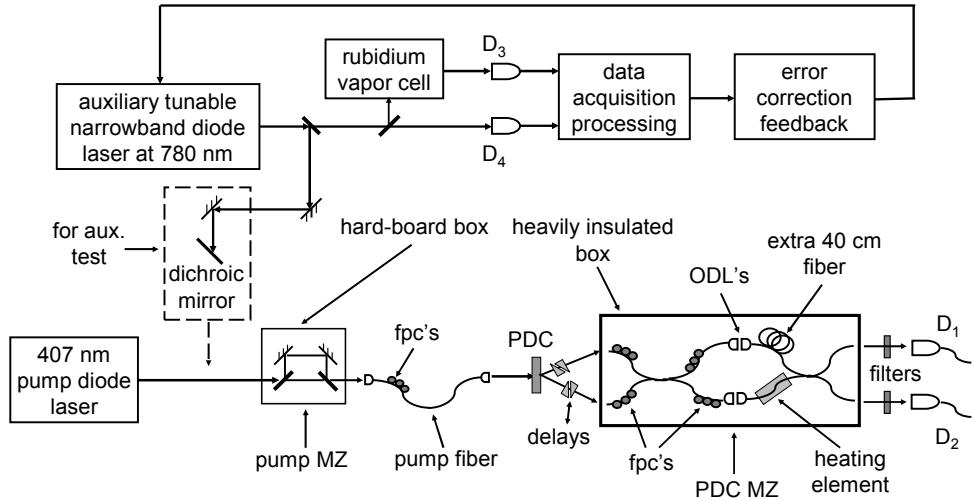}
\vspace{-0.2in} \caption{Experimental apparatus used to implement
the conceptual diagram of Figure 1. Here the pump Mach-Zehnder
interferometer is realized using bulk free-space optics, while the
PDC Mach-Zehnder interferometer is comprised of single-mode fiber
components. The upper part of the figure shows an auxiliary
frequency-stabilized narrowband laser system that is used to
calibrate and test the stability and phase-scanning of the pump and
PDC interferometers. The PDC source (a BBO crystal) is pumped by a
short coherence-length ultra-violet diode laser at 407 nm. The
experimental studies consist of monitoring the PDC coincidence
counting rate between two single photon detectors $D_{1}$ and
$D_{2}$. Additional details and symbols are found in the main text.}
\end{figure*}

It is instructive to use equation (\ref{eq:4terms}) to compare this
intermediate regime of two-photon interferometry using a short
coherence-length CW pump, with the extreme cases of Gisin's time-bin
method using a short pumping pulse and traditional Franson
interferometry with long coherence-length CW pumps. In the time-bin
case, the use of a single well-defined short pump pulse provides a
``starting clock'' for the experiments \cite{brendel99}. This
provides timing information which allows the third and fourth terms
in equation (\ref{eq:4terms}) to be easily discarded based on the
detector firing times. The remaining first and second terms in
equation (\ref{eq:4terms}) are indistinguishable in time, and share
a common pump-phase regardless of the coherence properties of the
pump pulse. In this way, the use of a pump MZ interferometer
cleverly induces the required second-order coherence, and 100\%
visibility two-photon interference effects can be observed with a
short pulsed pump \cite{brendel99,deriedmatten04,marcikic04}.

On the other extreme,  Franson-type interferometry without a pump MZ
results in only two terms, rather than the four listed in equation
(\ref{eq:4terms}). These two terms correspond to both photons taking
the long paths (eg. $|L_{s}L_{i}\rangle $) or both taking the short
paths (eg. $|S_{s}S_{i}\rangle $) through the PDC MZ interferometer.
In this case the long-long term has the relevant pump phase
$\phi_{o}(t^{\prime}-\tau)$, while the short-short term corresponds
to the initial pump phase $\phi_{o}(t^{\prime})$. However, by using
a CW pump with a coherence length much greater than $c\tau$, these
two terms remain coherent, and 100\% visibility two-photon
interference effects can be observed \cite{franson89}. In this case,
the required second-order coherence originates from the pump laser
itself.

In the intermediate regime demonstrated here, the lack of
second-order coherence due to the short coherence length of the pump
is overcome to some extent by the addition of the pump MZ
interferometer.  For any point in time, the use of the long and
short arms in the pump MZ provide two distinct points in the CW pump
beam that share a common value of initial pump phase $\phi_{o}(t)$.
Matching the separation $\tau_{p}$ of these two points to the delay
$\tau$ in the PDC MZ results in the first two indistinguishable
two-photon amplitudes in equation (\ref{eq:4terms}), in exact
analogy to the pulsed time-bin case.  However, the CW nature of the
pump precludes the use of a ``starting clock'' to cut-off the
remaining two terms which provide an incoherent background.

Under otherwise ideal experimental conditions, this incoherent
background limits the two-photon interference visibility to 50\%. It
can be shown that the four-term state in equation (\ref{eq:4terms})
leads to a coincidence counting rate given by:

\begin{equation}
R_{c}= 1 - \frac{1}{2}Cos(\phi_{p} - \phi)
\label{eq:ccr}
\end{equation}

\noindent where the factor of $1/2$ is the visibility. Because the
phase shift $\phi$ corresponds to the N00N state deBroglie
wavelength \cite{edamatsu02}, the coincidence counting rate is
expected to show oscillations at the pump wavelength as a function
of the relative path lengths in the PDC MZ interferometer.

\section{Experimental setup}
\label{sec:experiment}

Figure 2 shows a schematic diagram of the apparatus we used to
implement the scheme of Figure 1. A CW ultraviolet diode laser beam
($\sim$40 mW at 407 nm, 0.1 nm linewidth) was sent through a
free-space pump MZ comprised of two 50/50 beamsplitters and two
mirrors. The pump MZ had a path length imbalance of approximately
$c\tau_{p}$ = 60 cm. The beam coming out of one outport port of the
pump MZ was coupled into a single-mode pump fiber. The use of a
single-mode pump fiber facilitated alignment of the free-space pump
MZ interferometer and provided a convenient and robust way to
deliver the pump beam to the PDC crystal. Due to losses in the pump
MZ, and a non-ideal transverse mode-shape of the diode laser pump,
the output of this fiber was typically $\sim$15 mW. This free-space
output was then focused into a 0.7 mm thick BBO crystal with
phase-matching considerations oriented for type-I degenerate PDC.

The PDC source produced pairs of horizontally polarized photons at
814 nm, which were coupled into a single-mode 3dB fiber coupler (eg.
a 50/50 fiber beamsplitter) that formed the front-half of a
fiber-based PDC MZ interferometer. Each arm of the PDC MZ contained
a fiber-coupled adjustable air-gap optical delay line (OZ Optics,
Model ODL 200). These delay lines had a travel range of 25 mm, and
could be used to adjust the relative path lengths in the device.

The back-half of the PDC MZ was a formed by a second 3dB fiber
coupler. The outputs of the PDC MZ were coupled back into free
space, and passed through interference filters with a 10 nm FWHM
centered at 814 nm. Two single photon detectors and standard
coincidence-counting electronics were then used to record the data.
Paddle-wheel style fiber polarization controllers (fpc's) were used
in several locations to control the polarizations in the
experiments. The entire fiber-based PDC MZ interferometer was built
on a large 0.5 inch thick aluminum plate contained within a heavily
insulated box to provide passive thermal stability and minimize
phase-drift.

In order to match the 60 cm free-space path length imbalance of the
pump MZ, an extra $\sim$40 cm of fiber was added to the long-arm of
the fiber-based PDC MZ interferometer. To ensure that $\tau$ exactly
matched $\tau_{p}$, we temporarily connected the pump MZ directly in
series with the PDC MZ and used an auxiliary broadband
``white-light'' source as the input. Small adjustments to the
air-gap delay lines were then made to find first-order interference
and match the optical delays in the two interferometers.

\section{Pump MZ stability tests}
\label{sec:stability test}

As can be seen from equation (\ref{eq:ccr}), the coincidence
counting rate depends on both $\phi$ and $\phi_{p}$. Because we are
interested in two-photon interference effects in the PDC MZ, it was
necessary to ensure that $\phi_{p}$ was held stable while $\phi$ was
scanned in a controlled way. We therefore conducted a number of
auxiliary tests to verify the stability of the pump MZ and the
ability to scan $\phi$.

In practice, we passively scanned $\phi$ using a thermally-induced
phase drift. As shown in Figure 2, a resistive heating element was
used on a small segment ($\sim$ 5 cm) of fiber in the short-arm of
the PDC MZ. We developed a procedure in which the heating element's
temperature was raised 4$^{o}$ and then quickly turned off. After
approximately 20 minutes into the subsequent cooling process, the
PDC MZ interferometer drifted fairly linearly as a function of time,
and went through roughly three 407 nm deBroglie periods in about 15
minutes. The entire free-space pump MZ interferometer was kept
inside a simple non-insulated hardboard box to minimize air-current
and thermal disturbances. This kept the pump interferometer
relatively stable, and there was negligible phase drift in
$\phi_{p}$ over a time period much longer than the 15 minutes needed
to scan $\phi$ through several fringes.

\begin{figure}
\includegraphics[width=3.5in]{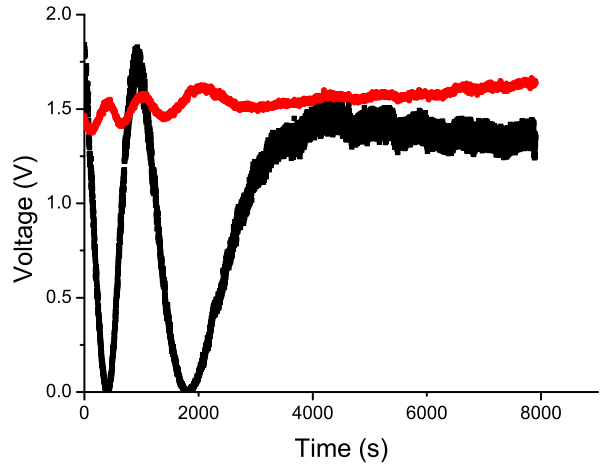}
\caption{ Typical results of a stability test of the pump
Mach-Zehnder interferometer. The long coherence-length auxiliary
laser at 780 nm and the pump laser at 407 nm are simultaneously sent
through the pump interferometer. Detectors monitor their outputs
after an intentional large thermal disturbance to the
interferometer. The auxiliary 780 nm signal (black trace)
demonstrates the required stability of the pump interferometer phase
during the experiments of interest, while the 407 nm signal (red
trace) demonstrates a lack of significant first-order interference
for the pump laser, as desired.}
\end{figure}

In order to verify the long-term stability of the pump MZ,  we used
the auxiliary narrowband external-cavity diode laser setup shown in
the top part of Figure 2. The same laser setup could also be used to
roughly calibrate the heating procedure used for scanning $\phi$ in
the PDC MZ interferometer.  The laser (New Focus Inc. Velocity Model
6312) had a linewidth $<$ 300 KHz, and a coherence length that
greatly exceeded the optical path length imbalance of both the pump
and PDC MZ interferometers. As shown in Figure 2, a standard
feedback loop based on an absorption measurement in a Rubidium vapor
cell was used to stabilize the wavelength of the laser. We typically
locked the laser's frequency to within 20 MHz of one of the ground
state hyperfine levels of the $D_{2}$ transition near 780 nm
\cite{ye96}. The locking was performed every 25 ms, which was
frequent enough to prevent this particular laser from drifting more
than $\sim 10^{-6}$ wavelengths from a fixed value. This locking
procedure kept any phase-noise due to wavelength drift to a
negligible level during our tests.

Figure 3 shows results from a typical stability test based on
first-order interference measurements in the pump MZ interferometer.
A dichroic mirror was temporarily added to inject both the 780 nm
laser, as well as the 407 nm pump laser, into the pump MZ. The 780
nm light was monitored by a detector in the lower output port of the
pump MZ, while the 407 nm light was simultaneously monitored by a
detector placed in the output of the single-mode pump fiber. To
demonstrate the stability, the data shown in Figure 3 was recorded
just after an intentional disturbance to the pump MZ with a heat
gun. The 780nm signal (black trace) shows nearly 100\% first-order
interference fringes that die out as the effects of the disturbance
wear off. After $\sim$ 1 hour, the phase drift is seen to be
negligible. Multiple data runs under slightly different conditions
showed the same type of behavior, with minimal phase drifts after
the interferometer had stabilized. During our main experiments, we
typically started collecting PDC data after the pump MZ box had been
sealed for many hours, from which we expect a negligible $\phi_{p}$
phase drift over a 15 minute data collection period.

The second data set (red trace) shown in Figure 3 corresponds to the
407 nm pump signal collected during the same time interval as the
780 nm auxiliary laser signal (black trace). The key point here was
to demonstrate the minimal amount of first-order coherence displayed
by the pump laser, as desired for our experiments. We ensured that
the polarizations of the pump laser from both arms of the
interferometer were the same, and measured the contribution of the
total power coupled into the single-mode pump fiber to be roughly
60\% from the short arm, and 40\% from the long arm (this deviated
from the ideal 50/50 split due minor deviations in beamsplitter
reflectivities and fiber-coupling efficiencies). These conditions
predict a maximal visibility of 98\%  for a long coherence length
source, whereas the visibility of the red trace in Figure 3 is only
roughly 8\% for our short coherence-length 407 nm diode pump laser.

Although the 60 cm path length imbalance of the pump MZ greatly
exceeds the expected coherence length of the pump, it is well-known
that the output of a CW multimode diode laser can exhibit multiple
``peaks'' separated by a distance 2L, where L is the length of the
laser cavity \cite{goodmanbook,baek07}. In fact, two-photon
interference experiments have been performed where the interference
originates from the coherence between these types of recurring peaks
\cite{burlakov01,kwon09}. In contrast, the two-photon interference
effects of interest in our setup should be due to second-order
coherence induced by the pump MZ, rather than the laser itself
\cite{brendel99}. Consequently, we needed to verify that the 407 nm
pump laser exhibited the kind of minimal first-order interference
seen in the red trace of Figure 3.

We also performed an auxiliary Michelson interferometer test
\cite{baek07} on our 407 nm laser and found the spacing between the
recurring diode laser peaks to be ($4705.5 \pm 0.9$) $\mu$m. The
path length imbalance of the pump MZ therefore corresponds to a
separation on the order of 100 peaks.  Coherence between these
widely spaced peaks is possible, but unlikely to be large due to the
finite linewidths of the individual modes in the diode laser
\cite{rodriguez87}. At the same time, the $\sim$60 cm path length
imbalance was not intentionally set to exactly match a whole number
of peaks. We believe that a combination of these two effects nearly
eliminated the first-order pump interference in the red trace of
Figure 3, as desired.

\section{Experimental Results}
\label{sec:results}

\begin{figure}
\includegraphics[width=3.5in]{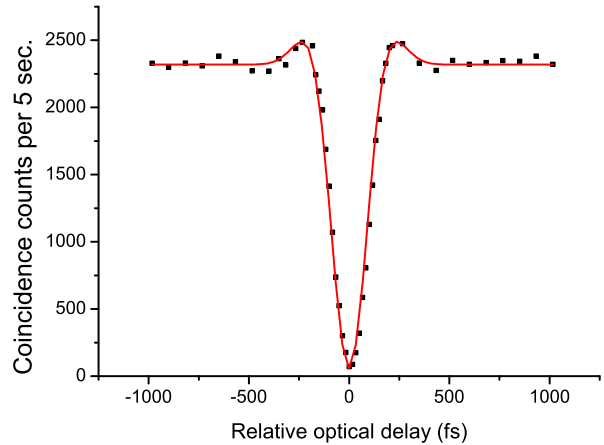}
\caption{ Results from the first step of the experimental procedure
showing the expected Hong-Ou-Mandel dip \cite{hong87,shih88} with a
($97.8 \pm 0.4)$\% visibility. This indicates a high-fidelity N00N
state in the output ports, as desired. The small shoulders in the
data and fit are due to the fact that the 10 nm bandpass filters
used in front of the detectors have a transmission profile that
slightly deviates from a Gaussian shape. }
\end{figure}

The results of our study can be best understood by considering four
sequential experimental steps that were used to build up to the
final conceptual diagram of Figure 1. The results of these four
steps are shown in Figures 4 - 7.

In the first step, the HOM interference effects which generate the
two-photon N00N state were optimized. To do this, the pump beam was
sent directly into the pump fiber (and then the PDC crystal) without
passing through a pump MZ interferometer. Additionally, the output
ports of the first PDC 50/50 beamsplitter were connected directly to
the detection channels, rather than serving as the inputs to the PDC
MZ interferometer. In this way, a standard fiber-based HOM
interferometer was temporarily formed \cite{hong87}. In our setup,
the relative delays between the two photons impinging on the HOM
beamsplitter were controlled with translating glass wedges in the
free-space paths after the PDC crystal, and FPC's on the input leads
of the beamsplitter were used to ensure the polarization states of
the PDC photons were the same.

Figure 4 shows the results of a typical HOM test in our setup. The
data shows a plot of the coincidence count rate as a function of the
relative optical delay between the two incident photons, with the
expected HOM ``dip'' at zero time delay. The data is best fit by a
dip-function with a visibility ($97.8 \pm 0.4$)\% which signifies
that both photons nearly always exit through the same output port of
the beamsplitter. This ensures a high-fidelity two-photon N00N
state, which is crucial for our subsequent experiments. Any
contamination by terms corresponding to one photon in each output
path of this first beamsplitter would lead to unwanted
$|L_{s}S_{i}\rangle $ amplitudes which are assumed to be
non-existent in our analysis of the full PDC MZ interferometer.

In the second step of the procedure, the outputs of the first 50/50
beamsplitter were reconnected as shown in Figure 2 to form the full
PDC MZ interferometer. However, the extra 40 cm long fiber was not
yet installed in the upper arm, so the two arms of the PDC MZ are
exactly balanced. Again, the pump MZ interferometer was not used. In
this configuration, our setup essentially reproduced the original
two-photon MZ interference experiment of Rarity et.al.
\cite{rarity90}. Although we are using a short coherence length CW
pump, the fact that the PDC MZ is balanced allows 100\% visibility
deBroglie wavelength fringes to be observed in the coincidence
counting rate \cite{rarity90}. This test allowed us to verify our
phase-scanning procedure and estimate the limitations on the
visibility due to mode-matching and other technical issues in our
apparatus.

\begin{figure}
\includegraphics[width=3.5in]{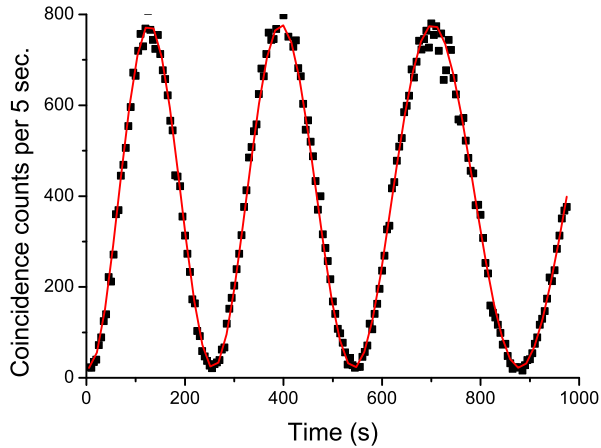}
\caption{Results from the second step of the experimental procedure
showing two-photon interference  with a balanced PDC MZ
interferometer and no pump MZ interferometer. In this step, the
arrangement is functionally equivalent to the well-known Rarity
interferometer \protect{\cite{rarity90}}. The data shows a
sinusoidal variation in the coincidence counting rate as a function
of time during a thermally-induced passive phase-scan, with a
best-fit visibility of ($93.0 \pm 0.7)$\%.}
\end{figure}

Figure 5 shows the results of this test when the relative phase in
the balanced PDC MZ interferometer was passively scanned using the
fiber-heating procedure described in Section \ref{sec:stability
test}. The data shows the expected sinusoidal variation in the
coincidence counting rate as a function of time. The maximum value
points correspond to relative phases differences that are even
multiples of $2\pi$, where constructive two-photon interference
deterministically splits the two-photon N00N state into exactly one
photon in each output port of the PDC MZ. This is simply the
time-reverse of the HOM effect \cite{pittman07,chen08}.  The slight
``stretching'' of the sinusoidal data in Figure 5 is due to the fact
that the relative phase drift is not exactly linear with time, as
would be expected from passive cooling of the heated-fiber segment
during this time interval. The data is best fit by a sinusoidal
function that takes this stretching into account, with a visibility
of ($93.0 \pm 0.7$)\%. This high visibility indicates a good degree
of indistinguishability between the interfering two-photon
amplitudes, and that phase-instabilities during each 5 second
coincidence-collection point are not a significant source of error
in our setup.

\begin{figure}
\includegraphics[width=3.5in]{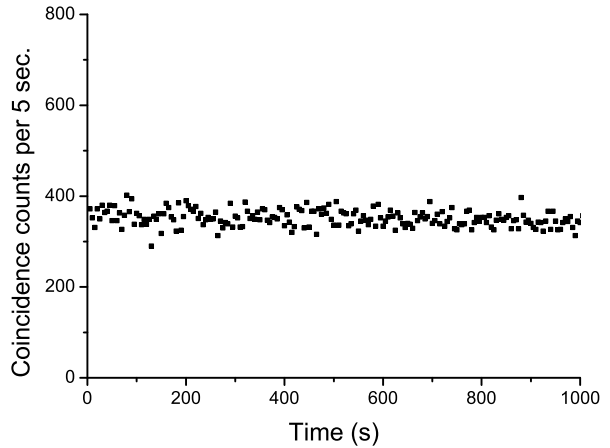}
\caption{Repeat of the data scan in Figure 5, but with the PDC MZ
interferometer path length difference intentionally set to be much
larger than the coherence length of the pump laser. Under these
unbalanced conditions, the two-photon interference effects seen in
Figure 5 are completely washed out, as expected.}
\end{figure}

The third step of the procedure involved repeating the same type of
scan shown in Figure 5, but with the extra 40 cm segment of fiber
installed in the upper arm of the PDC interferometer. In this case,
the path length imbalance of the PDC interferometer greatly exceeds
the coherence length of the pump laser. Consequently, there is no
coherence between the $|S_{s}S_{i}\rangle $ and $|L_{s}L_{i}\rangle
$ amplitudes, and no two-photon interference is expected
\cite{rarity90,franson89}. Figure 6 shows a result of the data
accumulated in this configuration under the same scanning procedure
used for Figure 5. It can be easily seen that the high-visibility
two-photon interference fringes of Figure 5 have completely vanished
due to the short coherence-length of the pump laser.

\begin{figure}
\includegraphics[width=3.5in]{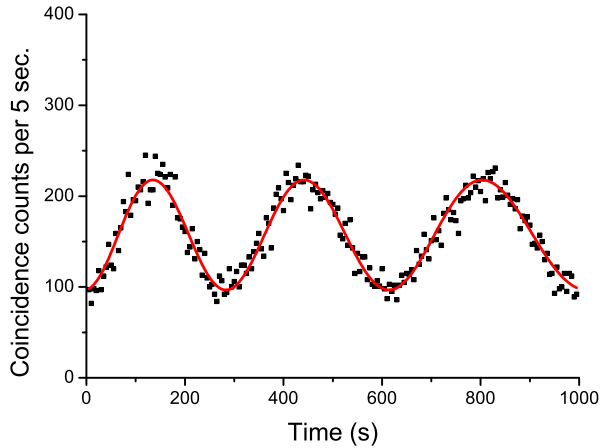}
\caption{Primary experimental results obtained by using the pump MZ
interferometer with its path length imbalance set to exactly match
that of the PDC MZ interferometer. This represents data taken with
the complete setup shown in Figure 2. As described in the discussion
of Figure 1, the inclusion of the pump MZ allows a recovery of
two-photon interference with a theoretical maximum visibility of
50\%.  The experimental data is fit by a sinusoidal function with a
($40.8 \pm 0.5)$\% visibility. }
\end{figure}

The fourth and final step is to partially recover these lost
two-photon interference effects by passing the pumping laser through
the free-space pump MZ interferometer shown in Figure 2, with its
path length imbalance set equal to the path length imbalance of the
PDC MZ (eg. $\tau_{p} = \tau$). As described in section
\ref{sec:overview}, the use of the pump MZ divides the two
interfering two-photon amplitudes into the four indistinguishable
amplitudes of equation (\ref{eq:4terms}). Two of these four
amplitudes are coherent in analogy with pulsed time-bin experiments
\cite{brendel99}, while the two remaining incoherent amplitudes can
not be cut-off due to the CW nature of the pump. This allows a
partial recovery of two-photon interference, with a theoretical
maximum visibility of 50\% given by equation (\ref{eq:ccr}).

Figure 7 shows the results of this final step when the relative
phase $\phi_{p}$ of the pump MZ was stabilized, and the relative
phase $\phi$ of the PDC MZ was scanned by the same fiber-heating
procedure used in Figures 5 and 6.  The data shows the predicted
sinusoidally varying coincidence counting rate, with a best-fit
visibility of ($40.8 \pm 0.5$)\%. The recovered two-photon
interference fringes seen in Figure 7 represent the primary
experimental result of this paper.

Note that due to the lack of significant first-order interference in
the pump MZ interferometer, only half of the original pump laser
power is delivered to pump fiber (and then the PDC crystal), while
the other half is wasted through the lower output port of the second
50/50 beamsplitter in the pump MZ. Consequently, the average
coincidence counting rate in Figure 7 is roughly half of those seen
in Figures 5 and 6. Additionally, the apparent deBroglie wavelengths
in Figures 5 and 7 are slightly different because the heating
procedure used to implement a phase scan is not actively controlled,
and varies slightly from run to run.

Furthermore, the 93\% visibility of Figure 5 obtained with a
balanced Rarity MZ setup \cite{rarity90} should ideally translate to
a 46.5\% visibility in the data of Figure 7. We believe the
reduction to 40.8\% in the actual experiment is primarily due to
alignment sensitivity in the HOM effect and PDC MZ interferometer,
rather than any phase instability in the pump MZ interferometer.
Additionally, the non-ideal 60/40 splitting of the pump MZ
interferometer led to a visibility loss of only 1\% due to a
slightly unequal weighting of the interfering $|S_{p};
L_{s}L_{i}\rangle$ and $|L_{p}; S_{s}S_{i}\rangle$ amplitudes in
equation (\ref{eq:4terms}), and was not a major factor. Despite
these issues, the primary results shown in Figure 7 clearly
demonstrate the expected interference fringes that can be obtained
in this intermediate regime of two-photon interferometry with a
short coherence-length CW pump laser.

The maximum expected two-photon interference fringe visibility of
only 50\%, and the use of only a single common PDC MZ interferometer
(as opposed to two spatially separated PDC MZ interferometers
\cite{franson89,brendel99}), raises the valid question of classical
analogues that could simulate the results of Figure 7
\cite{ou90,kaltenbaek07,franson09}. Indeed a single balanced ``PDC
MZ'' interferometer fed by a variety of simple classical light
sources can produce 100\% visibility fringes in the product of
photodiode signals from its output ports. The fact that we are
essentially using two unbalanced (but matched) MZ interferometers in
series does not fundamentally change the situation, and opens the
possibility of classical white-light-type interference effects that
can produce results related to Figure 7.

\section{Summary and discussion}
\label{sec:summary}

In some sense, the interference effects observed in PDC-based
two-photon interferometry originate from the uncertainty in the
emission time of the detected photon pair. In the original
Franson-type experiments \cite{franson89}, the use of a narrowband
CW pump laser renders the photon pair emission time completely
uncertain, while in the Gisin-type pulsed time-bin experiments the
emission time is confined to two possible well-defined values
\cite{brendel99}.  In both cases, however, the fact that the
two-photon state is a coherent superposition of the possible
emission times allows high-visibility non-classical interference
effects to be observed.  In this paper, we have described an
experimental demonstration of an intermediate situation that shares
features of both of these extreme cases.

By passing a short coherence-length CW pump laser through an
unbalanced time-bin-like pump MZ interferometer, coherence is
created between two possible pair emission times with a well-defined
separation, but an unknown overall starting time. As in the pulsed
time-bin case, this coherence allows two-photon interference effects
to be seen in a subsequent unbalanced PDC MZ interferometer,
regardless of the coherence properties of the pump laser itself
\cite{brendel99}.  However, the CW nature of the pumping beam here
precludes the use of the convenient ``starting clock'' available in
the pulsed-pump case. This prevents the ability to cut-off
incoherent two-photon amplitudes which limits the visibility to 50\%
here, as opposed to the 100\% visibility that is possible in the
pulsed case \cite{brendel99}.

The observation of these effects was facilitated by using a
two-photon N00N state in a single fiber-based unbalanced PDC MZ
interferometer \cite{rarity90}. In comparison to more general
experiments using two spatially separated PDC MZ interferometers
\cite{franson89,brendel99}, this provided a relatively robust and
stable experimental platform for an explicit demonstration the role
of pump coherence in an intermediate regime of two-photon
interferometry.

We acknowledge useful discussions with our colleague J.D. Franson.
This work was supported in part by the National Science Foundation
under grant No. 0652560.

%%%%%%%%%%%%%%%%%%%%%%%%%%%%%%%

\end{document}